\documentclass[twocolumn,showpacs,prb,aps,superscriptaddress]{revtex4}
\usepackage{amsmath}
\usepackage{graphicx}
\usepackage{color}
\newcommand{\be}{\begin{equation}}
\newcommand{\ee}{\end{equation}}
\newcommand{\bea}{\begin{eqnarray}}
\newcommand{\eea}{\end{eqnarray}}
\newcommand{\bs}{\begin{split}}
\newcommand{\bse}{\begin{subequations}}
\newcommand{\ese}{\end{subequations}}
\newcommand{\cacoas}{${\rm CaCo_{1.86}As_2}$}
\newcommand{\cacoastwo}{${\rm CaCo_2As_2}$}

\begin{document}
\title{Crystal and magnetic structure of CaCo$_{1.86}$As$_2$ studied by x-ray and neutron diffraction}
\author{D. G. Quirinale}
\affiliation {Ames Laboratory, USDOE and Department of Physics and Astronomy, Iowa State University, Ames, Iowa 50011, USA}
\author{V. K. Anand}
\altaffiliation{Present address: Helmholtz-Zentrum Berlin f\"{u}r Materialen und Energie, Hahn-Meitner-Platz~1, 14109 Berlin, \mbox{Germany}.}
\affiliation {Ames Laboratory, USDOE and Department of Physics and Astronomy, Iowa State University, Ames, Iowa 50011, USA}
\author{M. G. Kim}
\affiliation {Ames Laboratory, USDOE and Department of Physics and Astronomy, Iowa State University, Ames, Iowa 50011, USA}
\affiliation{Materials Sciences Division, Lawrence Berkeley National Laboratory, Berkeley, California 94720, USA}
\author{Abhishek Pandey}
\affiliation {Ames Laboratory, USDOE and Department of Physics and Astronomy, Iowa State University, Ames, Iowa 50011, USA}
\author{A. Huq}
\affiliation{Spallation Neutron Source, Oak Ridge National Laboratory, Oak Ridge, TN, 37830, USA}
\author{P. W. Stephens}
\affiliation{Department of Physics and Astronomy, SUNY at Stony Brook, Stony Brook, New York 11974, USA}
\author{T. W. Heitmann}
\affiliation{The Missouri Research Reactor, University of Missouri, Columbia, Missouri 65211, USA}
\author{A. Kreyssig}
\author{R. J. McQueeney}
\author{D. C. Johnston}
\author{A. I. Goldman}
\affiliation {Ames Laboratory, USDOE and Department of Physics and Astronomy, Iowa State University, Ames, Iowa 50011, USA}

\date{\today}

\begin{abstract}

Neutron and x-ray diffraction measurements are presented for powders and single crystals of \cacoastwo. The crystal structure is a collapsed-tetragonal ${\rm ThCr_2Si_2}$-type structure as previously reported, but with 7(1)\% vacancies on the Co sites corresponding to the composition ${\rm CaCo_{1.86(2)}As_2}$.  The thermal expansion coefficients for both the $a$ and $c$~axes are positive from 10 to 300~K\@. Neutron diffraction measurements on single crystals demonstrate the onset of A-type collinear antiferromagnetic order below the N\'eel temperature $T_{\rm N} = 52$(1)~K with the ordered moments directed along the tetragonal $c$-axis, aligned ferromagnetically in the \emph{ab}-plane and antiferromagnetically stacked along the \emph{c}-axis.

\end{abstract}


\maketitle

\section{\label{Intro} INTRODUCTION}

The cobalt arsenides $A$Co$_2$As$_2$ ($A$ = Ca, Sr, Ba) with the tetragonal ThCr$_2$Si$_2$ structure (space group $I4/mmm$) have received increased attention recently because of their close relationship to the parent compounds of the $A$Fe$_2$As$_2$ superconductor family.\cite{Johnston_2010,PandG_2010,Canfield_2010,Stewart_2011}  The isostructural $A$Fe$_2$As$_2$ parent compounds manifest structural and magnetic transitions at approximately the same temperature, from a high-temperature paramagnetic tetragonal phase to a low-temperature orthorhombic antiferromagnetic (AFM) structure characterized by the magnetic propagation vector, \textbf{Q}$_{\rm{AFM}}$ = ($\frac{1}{2}, \frac{1}{2}, 1$), the so-called stripe-like magnetic phase.\cite{Huang_2008,Jesche_2008,Goldman_2008} Various substitutions can be made for any of the three different atomic sites, which reduce the structural and magnetic transition temperatures. Doping $A$Fe$_2$As$_2$ with K,\cite{Rotter_2008} Co,\cite{Sefat_2008,Ni_2008} Ni,\cite{Li_2009,Canfield_2009} Rh,\cite{Ni_2009,Han_2009} P,\cite{Kasahara_2010} Pd,\cite{Ni_2009,Han_2009} Ir,\cite{Han_2009} Pt,\cite{Saha_2010} or Ru\cite{Sharma_2010,Thaler_2010,Kim_2011} also induces superconductivity (SC) over a finite range in concentration.  For example, upon doping with Co for Fe in Ba(Fe$_{1-x}$Co$_x$)$_2$As$_2$, both the structural and AFM order are suppressed to lower temperatures and split, with the structural transition found at a slightly higher temperature than the magnetic transition.\cite{Ni_2008,Chu_2008,Lester_2009} SC emerges for Co substitutions between $x$ = 0.03 and $x$ = 0.12 with superconducting transition temperatures ($T_{\rm{c}}$) up to $\approx$23~K. Neutron diffraction studies have clearly established the coexistence of, and competition between, SC and AFM order over a small range in Co concentration.\cite{Pratt_2009a,Christianson_2009,Fernandes_2010,Nandi_2010}  Above $x$ $\approx$ 0.06, the AFM order is completely suppressed, but correlated AFM fluctuations are observed at \textbf{Q}$_{\rm{AFM}}$. Co substitution levels above $x$ $\approx$ 0.12, lead to the suppression of both the correlated AFM fluctuations and SC.\cite{Sefat_2008,Ni_2008,Matan_2009,Sato_2011,Leithe_2008}

CaFe$_2$As$_2$ \cite{Ni_2008a,PC_2009} presents a particularly interesting case.  At ambient pressure, the substitution of Co or Rh for Fe \cite{Kumar_2009,Matusiak_2010,Harnagea_2011,Ran_2012,Danura_2011} results in the suppression of AFM order and SC emerges with $T_{\rm{c}}$ up to $\approx$20~K. Under modest applied pressure CaFe$_2$As$_2$ manifests fascinating new behavior including a transition to an isostructural volume-collapsed tetragonal (cT) phase\cite{Kreyssig_2008,Goldman_2009} that is nonmagnetic\cite{Kreyssig_2008,Goldman_2009,Pratt_2009b,Soh_2013} and nonsuperconducting.\cite{Yu_2009} The cT phase in CaFe$_2$As$_2$ is distinguished by a striking 9.5\% reduction in the tetragonal \emph{c}-axis lattice parameter, with respect to the high-temperature ambient-pressure tetragonal (T) phase, along with the absence of the stripe-like magnetic order found for the low-temperature ambient-pressure orthorhombic phase and of the magnetic fluctuations found in the T phase.

In some respects the $A$Co$_2$As$_2$ compounds appear to exhibit very different behavior as compared to the iron arsenides.  BaCo$_2$As$_2$ and SrCo$_2$As$_2$ are metals with enhanced paramagnetic susceptibilities and neither manifest magnetic order or SC down to 2~K.\cite{Sefat_2009,Pandey_2013}  However, recent inelastic neutron scattering measurements have revealed that ${\rm SrCo_2As_2}$ features AFM fluctuations at the same wave vector, \textbf{Q}$_{\rm{AFM}}$ = ($\frac{1}{2}, \frac{1}{2}, 1$), ubiquitous in the $A$Fe$_2$As$_2$ parent compounds.\cite{Jayasekara_2013} This is surprising because, for SrCo$_2$As$_2$, there is no clear nesting feature at this wave  vector that would suggest an instabiltity toward magnetic ordering.\cite{Pandey_2013}

\cacoastwo\ is reported to have a cT ${\rm ThCr_2Si_2}$-type structure with a small $c/a = 2.59$.\cite{Pfisterer_1980,Pfisterer_1983} Two groups have independently reported the bulk magnetic and transport properties of single crystals of this compound.\cite{Cheng_2012, Ying_2012} In contrast to the nonmagnetic state that characterizes the cT phase of CaFe$_2$As$_2$, Cheng et al.\ found AFM order in ${\rm CaCo_2As_2}$ below $T_{\rm N} = 76$~K with two successive magnetic field-induced transitions at $H = 3.5$~T and 4.7~T with a narrow hysteresis.\cite{Cheng_2012}  On the other hand Ying et al.\ observed AFM order below $T_{\rm N} = 70$~K and only one spin-flop transition at 3.5~T. \cite{Ying_2012} Ying et al.\ also reported that a 10\% Sr substitution for Ca increases $T_{\rm N}$ from 70~K to 90~K whereas the spin-flop field decreases from 3.5~T in ${\rm CaCo_2As_2}$ to 1.5~T in ${\rm Ca_{0.9}Sr_{0.1}Co_2As_2}$.\cite{Ying_2012} Both groups grew their single crystals using the CoAs self-flux growth technique and suggested that the AFM structure is collinear A-type, in which the Co moments within the $ab$-plane are ferromagnetically aligned along the $c$-axis and the moments in adjacent layers are aligned antiferromagnetically.  However, the proposed A-type AFM order has not previously been confirmed via diffraction techniques and is surprising in light of the inelastic neutron scattering measurements in SrCo$_2$As$_2$ mentioned above which show stripe-like AFM fluctuations.\cite{Jayasekara_2013}

Here we report on our crystal and magnetic structure investigations of powders and single crystals of Sn-flux grown \cacoastwo\ by x-ray and neutron diffraction.  The co-refined x-ray and neutron powder diffraction measurements confirm the essential elements of the structure determined previously.\cite{Pfisterer_1980,Pfisterer_1983} However, from two independent measurements, we show that there are 7(1)\% vacancies on the Co site, leading to a stoichiomentry of ${\rm CaCo_{1.86(2)}As_2}$ instead of the previously reported stoichiometry \cacoastwo. We also report on the temperature dependence of the lattice constants which exhibit conventional behavior in contrast to the negative thermal expansion found previously for the \emph{c}-axis of SrCo$_2$As$_2$.\cite{Pandey_2013} Our single-crystal magnetic neutron diffraction measurements are consistent with collinear A-type AFM order, as suggested by Cheng \emph{et al} and Ying \emph{et al}.\cite{Cheng_2012, Ying_2012} The $T_{\rm N} = 52(1)$~K that we consistently observe is about 20~K lower than the recent reports\cite{Cheng_2012, Ying_2012} for, as yet, unknown reasons.

\section{\label{ExpDetails} Experimental Details}

Single crystals of \cacoas\ were prepared by solution growth using Sn flux. First, a polycrystalline sample was synthesized by the standard solid state reaction route starting with high purity Ca (99.98\%), Co (99.998\%) and As (99.99999\%) from Alfa Aesar in the stoichiometric 1:2:2 ratio. The polycrystalline ${\rm CaCo_2As_2}$ and the Sn flux (99.999\%, Alfa Aesar) were taken in a 1:20 molar ratio, placed in an alumina crucible, and sealed inside an evacuated silica tube. The sealed tube was then slowly heated to 1150~$^\circ$C at a rate of 100~$^\circ$C/h and soaked for 30~h. The crystal growth was carried out by slow-cooling the solution to 800~$^\circ$C at a rate of 3~$^\circ$C/h. Shiny plate-like crystals of typical size $3 \times 3 \times 0.4$~mm$^3$ were obtained by centrifuging the flux at 800~$^\circ$C.

A JEOL JXA-8200 electron probe microanalyzer was used to determine the chemical composition of the crystals using wavelength-dispersive x-ray spectroscopy (WDS) analysis. A Quantum Design, Inc., superconducting quantum interference device magnetic properties measurement system (MPMS) was used for the magnetic susceptibility, $\chi(T)$, measurements.

High-resolution x-ray powder diffraction data were collected at ambient temperature on beamline X16C at the National Synchrotron Light Source.  The crushed-crystal samples were loaded into a 1~mm diameter glass capillary along with Si powder which served as a standard for the refinement of the lattice parameters. The x-ray wavelength, 0.6995~\AA, was chosen using a Si(111) double monochromator. The powder diffraction pattern was collected in the 5--45$^\circ$ $2\theta$ range with a constant step size of 0.005$^\circ$, and a linearly varying counting time of 1--3~s/point. The incident beam intensity was monitored with an ion chamber and the diffracted radiation was measured with a NaI scintillation detector. The axial and in-plane resolution of the diffractometer were set by slits and a Ge(111) analyzer crystal, respectively. Diffraction data were obtained at temperatures of 295~K, 200~K, 100~K, 60~K and 20~K\@.

Neutron powder diffraction data were collected at the Spallation Neutron Source (POWGEN) at Oak Ridge National Laboratory.  A sample of mass 1.3~g was prepared by crushing single crystals and loaded into a vanadium sample can in the presence of He exchange gas to maintain thermal conductivity at low temperatures.  Data were collected at temperatures of 300~K and 10~K using center wavelengths of 1.333 and 4.797~\AA\ which allowed the acquisition of diffraction patterns for $d$ spacings of 0.45 to 8.5~\AA\@.  To search for evidence of magnetic order at low temperature, the longer-wavelength data at 10~K were collected for six times longer than for 300~K as the AFM signal was expected to be weak because of the expected small ordered moment ($\approx$ 0.3~$\mu_{\rm{B}}$/Co).\cite{Cheng_2012,Anand_2013}

Single-crystal neutron diffraction measurements were performed on a 34~mg crystal using the TRIAX triple-axis spectrometer at the University of Missouri Research Reactor.  The horizontal beam collimation before the monochromator, between the monochromator and sample, between the sample and analyzer, and between the analyzer and detector were 60$^\prime$-40$^\prime$-40$^\prime$-80$^\prime$, respectively. Incident and scattered neutrons with an energy of 14.7~meV were selected by a pyrolitic graphite (PG) (002) monochromator and PG (002) analyzer, and two PG filters were employed to effectively eliminate higher harmonics in the incident beam. The sample was aligned such that the ($h$~0~$\ell$) reciprocal lattice plane was coincident with the scattering plane of the spectrometer, and was mounted on the cold finger of a helium closed-cycle refrigerator.

\section {Results}

\subsection{Structure Determination from Powder X-ray and Neutron Diffraction Measurements}

\begin{figure}
\includegraphics[width=1\linewidth]{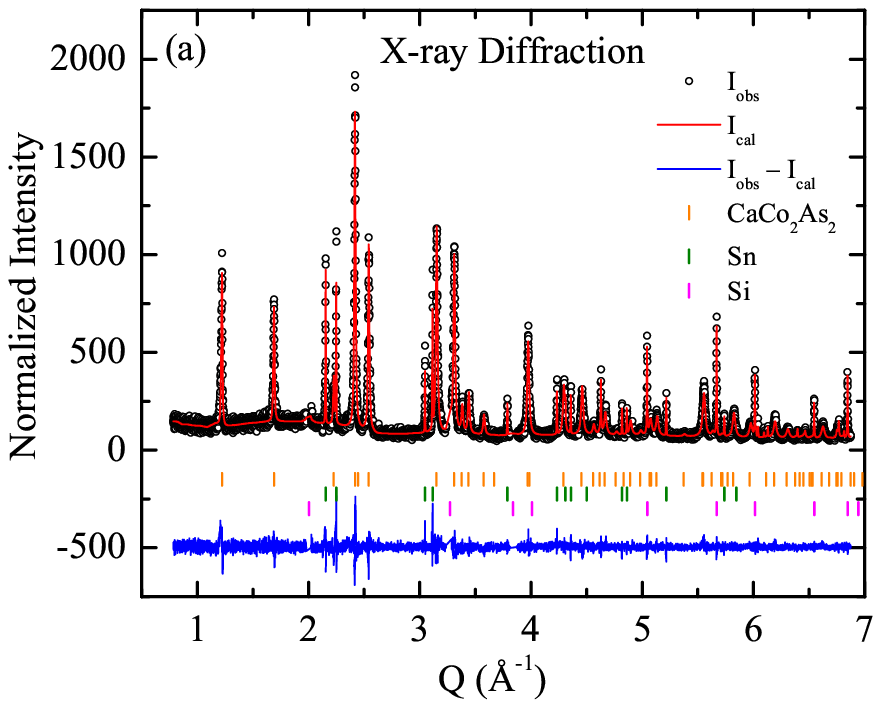}
\includegraphics[width=1\linewidth]{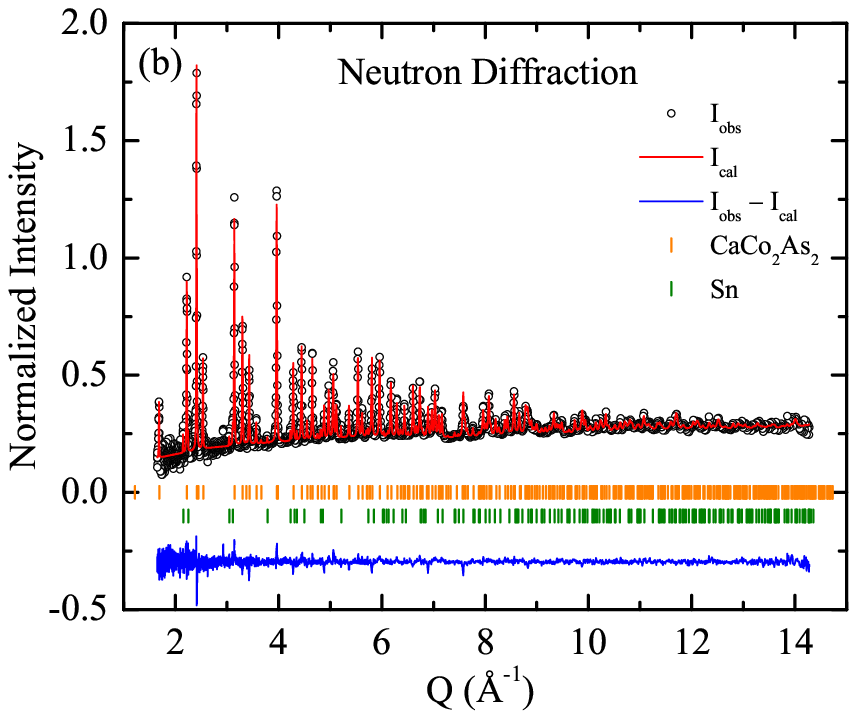}
\caption{(Color online) (a) X-ray powder diffraction and (b) neutron powder diffraction profile fits from Rietveld analysis of the \cacoastwo\ samples at 300 K.  The data in (a) contain peaks of the Si powder used as an internal lattice parameter standard.  Both patterns contain peaks from the Sn flux. Shown are the measured data (open circles), the calculated intensity pattern (red lines), and the corresponding residuals (i.e. the difference between observed and calculated pattern) by the curve below the fit. The vertical lines indicate the positions of diffraction peaks corresponding to phases used in the refinements.}
\label{Fig:xrd_neut}
\end{figure}

Rietveld co-refinements of the ambient-temperature x-ray powder data and the neutron powder patterns at both 300~K and 10~K were done together using the General Structure Analysis System package (GSAS)\cite{Larson_2004} and the graphical user interface (EXPGUI).\cite{Toby_2001} The 300~K neutron and x-ray powder data and refinements are shown in Fig.~\ref{Fig:xrd_neut} and the results of the co-refinement are given in Tables~\ref{Table:CrystalData} and \ref{tab:XRD2}. For the neutron data, the absorption parameter at 10~K was kept to the same value as obtained from the 300~K data as this is highly correlated to the thermal parameters and is not expected to change as a function of temperature. For the x-ray data, the strongest Si lines at low Q were excluded from the refinements.

\begin{figure}
\centering\includegraphics[width=0.6\linewidth]{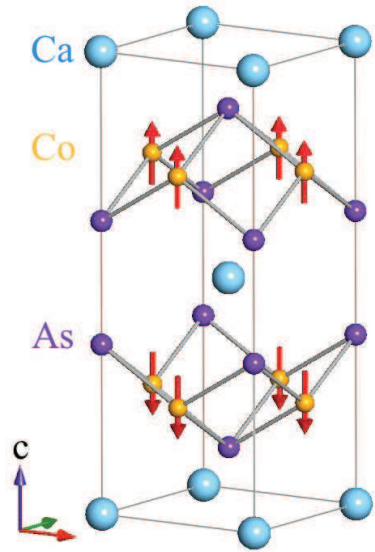}
\caption{(Color online) The body-centered tetragonal chemical and magnetic unit cell of \cacoas\ (space group $I4/mmm$).  The red arrows denote the ordered magnetic moment directions in the A-type antiferromagnetic structure determined from single-crystal neutron diffraction measurements.}
\label{Fig:structure}
\end{figure}

The refinement confirmed that \cacoas\ crystallizes in the ${\rm ThCr_2Si_2}$-type body-centered tetragonal structure (space group $I4/mmm$) shown in Fig.~\ref{Fig:structure}. The lattice parameters and the $z$-coordinate of the As atom are in good agreement with the literature values.\cite{Pfisterer_1980,Pfisterer_1983} We obtain the $c/a$ ratio and the interlayer As--As distance $d_{\rm As-As} = (1-2z_{\rm As})c $ using the crystallographic data in Table~\ref{Table:CrystalData}, yielding $c/a=2.576$ and $d_{\rm As-As} = 2.753$~\AA\@. These values indicate that \cacoas\ is in the cT phase.  Indeed, the value of $d_{\rm As-As}$ and $c/a$ for \cacoas\ are close to the respective values observed for the pressure-induced cT phase of ${\rm CaFe_2As_2}$.\cite{Kreyssig_2008, Goldman_2009}

The major impurity phase found in the samples is the adventitious Sn flux which comprises approximately 7\% of the sample by weight, and was included in the refinement. CoAs, at a level of $< 2$~wt\%  was also noted in the x-ray data.  A few very low-intensity diffraction peaks in the x-ray pattern could not be identified conclusively with other known phases.  Both the x-ray and neutron refinements (together and separately) show that the compound is deficient in Co, giving a composition of ${\rm CaCo_{1.86(2)}As_2}$. WDS analysis was done on several crystals to obtain an independent estimate of the Co vacancy concentration. The average atomic ratios obtained from the WDS analysis are Ca\,:\,Co\,:\,As = 20.2(2)\,:\,38.3(3)\,:\,40.8(4) which indicates the stoichiomentry to be ${\rm CaCo_{1.88(2)}As_2}$, consistent with the value obtained from the refinement of the powder x-ray and neutron data. Although we see a significant amount of Sn impurity in the powder data due to the adventitious Sn flux, our WDS analysis showed that about 0.5~mol\% of Sn is also present in the bulk of the crystals.

\begin{table*}
\caption{\label{Table:CrystalData}  Results of the Rietveld co-refinements of the synchrotron x-ray powder diffraction measurements together with the neutron powder diffraction data at 300~K and 10~K of \cacoas. Refined values are given for the lattice parameters $a$ and $c$, the Co concentration $x$ and the arsenic $c$-axis positional parameter $z_{\rm As}$, together with the goodness of fit parameters $R_{\rm p}$ and $R_{\rm wp}$. The overall value of the goodness of fit, $\chi^2$, was 2.18.  A number in parentheses represents the standard deviation in the last digit of a quantity. }
\begin{ruledtabular}
\begin{tabular}{l c c c c c c }
Composition       &  $x$  &   $a$  (\AA)  &  $c$  (\AA)  & $z_{\rm As}$ & $R_{\rm p}$, $R_{\rm wp}$ (\%) & Remarks \\
\hline
CaCo$_x$As$_2$ & 1.86(2) & 3.9906(1) & 10.2798(2) & 0.3661(2) & 0.092, 0.113 & x-ray, $T = 300$ K \\
CaCo$_x$As$_2$ & 1.86(2) & 3.9906(1) & 10.2798(2) & 0.3661(2) & 0.064, 0.036  & neutron, $T = 300$ K \\
CaCo$_x$As$_2$ & 1.86(2)& 3.9846(1) & 10.2019(2) & 0.3664(2) & 0.075, 0.042 & neutron, $T = 10$ K \\
\end{tabular}
\end{ruledtabular}
\end{table*}

\begin{table}
\caption{\label{tab:XRD2} Atomic coordinates in the space group $I4/mmm$ obtained from the Rietveld co-refinements of x-ray and neutron powder data from \cacoas\ at $T$ = 300~K.}
\begin{ruledtabular}
\begin{tabular}{ccccccc}
  Atom & Wyckoff   &	 $x$ 	&	$y$	&	$z$	 & Fractional \\	
   	& position 		& 			&		& 		& occupancy\\
   	& 				& 			&		& 		& (\%)\\
\hline
    Ca & $2a$  	&	 0 	&	0	&   0		& 100  \\
    Co & $4d$	    &	 0 	&	1/2	&	1/4 	& 93(1)  \\
    As & $4e$ 	&	 0 	&   0 	&	0.3661(2) & 100 \\
\end{tabular}
\end{ruledtabular}
\end{table}

\begin{figure}
\centering\includegraphics[width=0.9\linewidth]{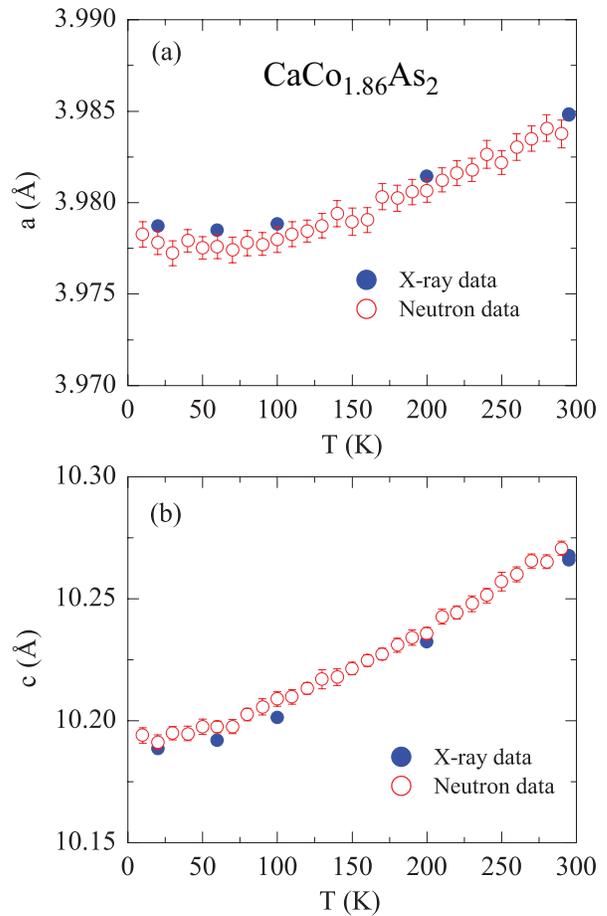}
\caption{(Color online)  Temperature dependence of the $a$ and $c$ lattice parameters for \cacoas\ determined by neutron single-crystal diffraction (open red circles) and x-ray powder diffraction (filled blue circles)}
\label{Fig:Lattice_Parameters}
\end{figure}

The temperature dependence of the $a$ and $c$ lattice parameters was determined from both single-crystal neutron and x-ray powder diffraction measurements, and the results are shown in Fig.~\ref{Fig:Lattice_Parameters}.  These results are conventional, but the $c$-axis data qualitatively differ from the temperature dependence of $c$ for ${\rm SrCo_2As_2}$, where $c$ {\it decreases} monotonically as $T$ increases from~7 to 300~K.\cite{Pandey_2013}  These x-ray and neutron data also show that $z_{\rm As}$ remains constant as temperature is varied from ambient down to 10~K\@.

\subsection{Magnetic Structure from Single-Crystal Neutron Diffraction}

\begin{figure}
\centering\includegraphics[width=0.8\linewidth]{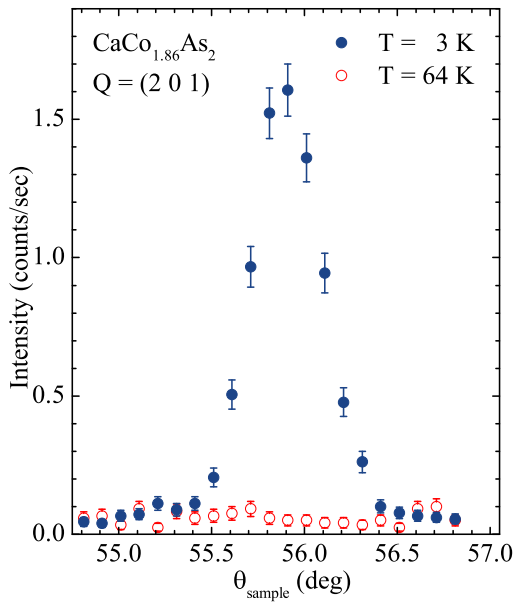}
\caption{(Color online) Sample-angle scan at the (2 0 1) magnetic Bragg peak position measured by neutron diffraction at $T$ = 3~K (filled blue circles), below $T_{\rm{N}}$ = 52(1)~K, and $T$ = 64~K (open red circles), above $T_{\rm{N}}$.}
\label{Fig:neutron_scan}
\end{figure}

\begin{figure}
\centering\includegraphics[width=0.8\linewidth]{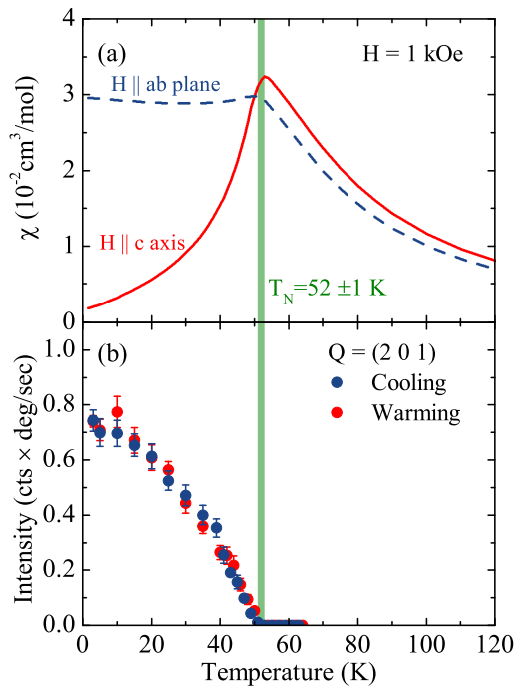}
\caption{(Color online)  Temperature evolution of (a) the magnetic susceptibility ($\chi$) for a field of 1 kOe parallel (solid line) and perpendicular (dashed line) to the tetragonal $c$-axis, and (b) the integrated intensity of the (2 0 1) magnetic Bragg peak on cooling (red filled circles) and warming (blue filled circles).  The vertical green bar denotes $T_{\rm{N}} = 52(1)~K$.}
\label{Fig:ordered_moment}
\end{figure}

We now turn to the single-crystal neutron diffraction determination of the magnetic order in \cacoas.  For temperatures above the N\'eel temperature $T_{\rm{N}}$ = 52(1)~K only Bragg peaks associated with the body-centered tetragonal chemical structure, with Miller indices $h + \ell = 2n$ (integer $n$) in the ($h~0~\ell$) scattering plane, were observed.  As shown in Fig.~\ref{Fig:neutron_scan}, below $T_{\rm{N}}$ new Bragg peaks appear at positions $h+\ell = 2n+1$, consistent with breaking the body-centered symmetry as expected for the predicted $A$-type antiferromagnetic (AFM) structure.\cite{Cheng_2012,Ying_2012}  Figure~\ref{Fig:ordered_moment}(b) plots the temperature evolution of the integrated intensity of the (2~0~1) magnetic Bragg peak as a function of temperature on cooling and warming.  No hysteresis was observed and the determined value $T_{\rm{N}}= 52(1) $~K is in excellent agreement with the value obtained from single-crystal $\chi(T)$ measurements shown in Fig.~\ref{Fig:ordered_moment}(a).

The direction of the ordered moment can be determined through a comparison of the measured intensities of magnetic Bragg peaks.  In particular, only the component of the ordered magnetic moment that is perpendicular to the scattering vector, \textbf{Q}$_{h k \ell}$, contributes to the magnetic peak with indices $(h~k~\ell)$. For \cacoas, below $T_{\rm{N}}$, we note that the ($0~0~\ell$) magnetic peaks with $\ell =$~odd are absent, demonstrating that the moments are directed along the tetragonal $c$-axis as illustrated in Fig.~\ref{Fig:structure}.

In principle, the magnitude of the magnetic moment can also be determined from a comparison of the magnetic and nuclear Bragg peak intensities.  However, the small mass of sample and its geometry (thin flat plate) allow us to place only an upper limit of $\sim~0.6~\mu_{\rm{B}}$/Co on the low-temperature ordered moment from these measurements, consistent with the small moment value of $\approx 0.3~\mu_{\rm{B}}$/Co estimated from the $M(H)$ measurements.\cite{Cheng_2012,Anand_2013}  We further note that our neutron powder diffraction measurements at 10~K failed to observe reflections at the magnetic Bragg peak positions, again consistent with a small ordered moment estimated from the single-crystal neutron diffraction experiment.

\section{Discussion}

In a cT compound, pnictogen (\emph{Pn}) bonding occurs between the \emph{Pn} atoms in adjacent transition metal-\emph{Pn} layers, where the \emph{Pn} atoms in adjacent layers are directly above and below each other (see Fig.~\ref{Fig:structure}).  This \emph{Pn} bonding effect was first observed in ${\rm CaCo_2P_2}$ in which the bonding between the P-P atoms results in a redistribution of electric charge and the oxidation state of P changes to P$^{-2} \equiv $ [P--P]$^{-4}/2$ from its formal oxidation state of P$^{-3}$ in the T phase.\cite{Hoffman_1985,Reehuis_1990,Reehuis_1998}

As shown previously, the pressure-induced transition from the T to the cT phase has been found to quench the Fe magnetic moment and associated AFM fluctuations in ${\rm CaFe_2As_2}$.\cite{Pratt_2009b,Soh_2013} In a chemical picture this can be viewed as a transition from the Fe$^{+2}$ oxidation state in the T phase to a Fe$^{+1}$ oxidation state in cT phase due to the appearance of As--As bonding. Thus there is a change in the formal oxidation state of the transition metal (TM) in a transition from the T to cT structure. Even though it is not clear whether the formation of a cT structure drives the change in oxidation state of the TM, or it is the stability of the oxidation state of TM that drives the formation of As-As bonding and the stability of the cT structure, it is quite evident from these observations that the formation of a cT structure and the formal oxidation state of TM are interrelated. In particular, a change in formal oxidation state of the TM may strongly modify the magnetic character of 122-type pnictides.

Co$^{+1}$ with a $3d^{8}$ electronic configuration is expected to be magnetic.  Indeed, there are several known cT 122-type Co compounds that exhibit interesting magnetic behavior due to the Co magnetism. ${\rm CaCo_2P_2}$ is one such example in which the Co atoms are in a magnetic $3d^8$ Co$^{+1}$ state. \cite{Reehuis_1990, Reehuis_1998} This compound exhibits A-type antiferromagnetism below $T_{\rm N}=113$~K, where the Co ordered moments are aligned ferromagnetically within the basal $ab$ plane and antiferromagnetically along the $c$-axis.\cite{Reehuis_1998} An interesting magnetic phase diagram driven by a T to cT transition with concurrent  P--P bonding has been observed for the system Sr$_{1-x}$Ca$_x$Co$_2$P$_2$. \cite{Jia_2009} High-temperature Co-sublattice AFM transitions have been found in cT rare earth compounds ${\rm CeCo_2P_2}$ ($T_{\rm N}=440$~K), ${\rm PrCo_2P_2}$ ($T_{\rm N}=304$~K), ${\rm NdCo_2P_2}$ ($T_{\rm N}=309$~K) and ${\rm SmCo_2P_2}$ ($T_{\rm N}=302$~K), whereas ${\rm LaCo_2P_2}$ exhibits ferromagnetic ordering with a Curie temperature $T_{\rm C}\approx 130$~K.\cite{Jeitschko_1985, Reehuis_1990, Reehuis_1998, Morsen_1988, Reehuis_1993, Reehuis_1994} The application of pressure has been found to result in Co-moment magnetic ordering in ${\rm EuCo_2P_2}$ that develops due to a pressure-induced isostructural phase transition from the T to the cT phase at a critical pressure $p_{\rm c} = 3.1$~GPa. \cite{Huhnt_1997,Chefki_1998} At ambient pressure, ${\rm EuCo_2P_2}$ orders antiferromagnetically below $T_{\rm N}=66.5$~K due to the ordering of localized Eu$^{+2}$ moments with spin $S = 7/2$ which is suppressed for pressures above $p_{\rm c}$ with a simultaneous development of itinerant Co-moment magnetic order at $T_{\rm N}=260$~K in the cT phase. \cite{Chefki_1998}

A comparison of the values of $c/a$ and $d_{\rm As-As}$ in the family of $AM_2X_2$ ($A$ = Ca, Sr, Ba; $M$ = Cr, Mn, Fe, Co, Ni, Cu, and $X$ = P, As) in Ref.~\onlinecite{Anand_2012a} demonstrates that \cacoas\ has a cT structure.  As a consequence of the formation of interlayer As--As bonds in the cT structure, the As atoms have an unusual oxidation state of As$^{-2} \equiv $ [As--As]$^{-4}$/2, whereas in the T structure of BaFe$_2$As$_2$ and ${\rm CaFe_2As_2}$, the formal oxidation state of As is As$^{-3}$. Therefore, the oxidation state of Co in cT-type \cacoas\ is Co$^{+1}$ whereas that of the Fe in the T phase of CaFe$_2$As$_2$ is Fe$^{+2}$.  Interestingly, our comparison of the magnetic properties of $AM_2X_2$ ($A$ = Ca, Sr, Ba; $M$ = $3d$ transition metal; $X$ = P, As) compounds in Ref.~\onlinecite{Anand_2012a} revealed that all the magnetically ordered compounds have the T structure except for the cT compound ${\rm CaCo_2P_2}$ which exhibits AFM order. This suggests that Co$^{+1}$ in the cT phase is magnetic in nature, consistent with the occurrence of AFM order in the cT-phase of ${\rm CaCo_2As_2}$.

\section{Summary}

Using both x-ray and neutron diffraction measurements on powders and single crystals of Sn-flux grown samples we have investigated the crystal and magnetic structures of \cacoas.  The co-refined x-ray and neutron powder measurements confirm the essential elements of the structure determined previously by Pfisterer and Nagorsen.\cite{Pfisterer_1980,Pfisterer_1983} However, we find, from both our powder diffraction and WDS measurements, that there are 7(1)\% vacancies on the Co site, leading to a stoichiomentry of ${\rm CaCo_{1.86(2)}As_2}$ instead of the previously reported stoichiometry of \cacoastwo.  We have also reported on the temperature dependence of the lattice constants which exhibit conventional behavior in contrast to the negative thermal expansion found previously for the \emph{c}-axis lattice parameter of SrCo$_2$As$_2$.\cite{Pandey_2013} Our single-crystal magnetic neutron diffraction measurements are consistent with collinear A-type AFM order, as suggested by Cheng \emph{et al} and Ying \emph{et al},\cite{Cheng_2012, Ying_2012} although the $T_{\rm N} = 52(1)$~K that we consistently observe is about 20 K lower than these reports.

\acknowledgments

This research was supported by the U.S. Department of Energy, Office of Basic Energy Sciences, Division of Materials Sciences and Engineering. Ames Laboratory is operated for the U.S. Department of Energy by Iowa State University under Contract No.~DE-AC02-07CH11358. Use of the National Synchrotron Light Source, Brookhaven National Laboratory, was supported by the U.S. Department of Energy, Office of Basic Energy Sciences, under contract No. DE-AC02-98CH10886. Research at ORNL's Spallation Neutron Source was sponsored by the Scientific User Facilities Division, Office of Basic Energy Sciences, U.S. Department of Energy.

\end{document}